\documentclass[fleqn,10pt]{wlscirep}
\usepackage[utf8]{inputenc}
\usepackage[T1]{fontenc}
\usepackage{graphicx}
\usepackage{hyperref}

\newcommand{\crit}{\ensuremath \mathrm{cr}}
\newcommand{\proton}{\ensuremath \mathrm{p}}
\newcommand{\Bor}{\ensuremath {{}^{11}\mathrm{B}}}
\usepackage{xcolor}

\title{Interaction of Laguerre-Gaussian laser pulses with borane targets of different hydrogen-boron ratio}

\author[1,2,*]{Lars Reichwein}
\author[2]{Alexander Pukhov}
\author[1,3]{Markus Büscher}
\affil[1]{Peter Grünberg Institut (PGI-6), Forschungszentrum Jülich, 52425 Jülich, Germany}
\affil[2]{Institut für Theoretische Physik I, Heinrich-Heine-Universität Düsseldorf, 40225 Düsseldorf, Germany}
\affil[3]{Institut für Laser- und Plasmaphysik, Heinrich-Heine-Universität Düsseldorf, 40225 Düsseldorf, Germany}
\affil[*]{l.reichwein@fz-juelich.de}

\begin{abstract}
We study the interaction of high-intensity Laguerre Gaussian laser pulses with hydrogen-boron compounds targets using 3D particle-in-cell simulations. The ratio of hydrogen to boron is varied throughout different simulation runs as a proxy model for various borane molecules that can be synthesized. We show that the strength of the axial magnetic fields generated via the Inverse Faraday effect depends on the specific ratio of target components, making boranes and the option to tune their composition of interest for proton-boron fusion. 
\end{abstract}
\begin{document}

\flushbottom
\maketitle

\thispagestyle{empty}

\section*{Introduction}

The recent breakthroughs in inertial confinement fusion (ICF) at the National Ignition Facility \cite{AbuShawareb2022, AbuShawareb2024} have led to intensified fusion research from both academic and commercial side.
Many of these approaches consider deuterium-tritium fusion as the cross-sections are comparatively large.
Besides this obvious choice for reactants, other reactions with smaller cross-sections can be of interest as well: in particular, for the case of proton-boron fusion,

\begin{equation}
   \proton + \Bor \to 3 \alpha \; ,
\end{equation}
no unwanted neutrons are produced which could otherwise have detrimental effects to reactor walls.
While achieving pB fusion on a commercial scale is significantly more complicated, nonetheless significant efforts are made in its research \cite{Batani2023, Molloy2025}. 

The approach to pB fusion can be in the form of pitcher-catcher style setups, where a pre-accelerated proton beam induces fusion reactions in a separate target \cite{Schollmeier2022,Liu2023}, or more direct in-target fusion \cite{Margarone2022, Istokskaia2023}. 
Thus, research on the topic of fusion has led to the investigation of many novel target structures, like foams \cite{Wei2023} which have been shown to be of relevance for several aspects of laser-matter interaction due to their highly localized high-density regions and effectively low density when homogenized by a laser pulse \cite{Gyrdymov2024}.

Beyond structured target approaches, the investigation of boron hydride compounds (``boranes'') has recently gained interest in the scope of nuclear fusion \cite{Krus2024, Turcu2024, Picciotto2024}.
These compounds contain the necessary reactants in a single molecule, theoretically making them a natural choice for in-target fusion.
Moreover, different borane compositions of various hydrogen-boron ratios can be synthesized \cite{Lipscomb2012}, which can influence the subsequent plasma dynamics during heating.

Kr\r{u}s \textit{et al.} have studied the prospect of pB fusion with boranes using the molecule \textit{anti-}B$_{18}$H$_{22}$ \cite{Krus2024} with the PALS laser at the Institute for Plasma Physics in Prague. The alpha-particle yield for these targets was observed to be in the range of $10^9 \,\mathrm{sr}^{-1}$.

Another aspect of fusion research is the generation of strong electromagnetic fields. It is known from several publications that circularly polarized Gaussian laser pulses can generate strong axial magnetic fields in underdense plasma via the Inverse Faraday Effect \cite{Haines2001, Kostyukov2001, Longman2021}. These generated fields could be utilized as confining fields in fusion-related setups or have also been proposed as possible pathways to study quantum electrodynamics with the next generation of laser facilities \cite{Samsonov2024}.

Other than Gaussian pulses, other structures carrying orbital angular momentum (OAM), like Laguerre-Gaussian (LG) laser pulses, can be utilized for the generation of axial fields as well.
The strength of the generated fields depends on the specific mode $\ell, p$ of the pulse as well as the target composition.

In particular, Longman and Fedosejevs showed\cite{Longman2021} that the axial field strength can be estimated as 
\begin{equation}
    B_x \propto |\psi_\ell|^2 \times \left[ \ell \left( \frac{|\ell| w_0^2}{r^2} - 2\right) + \sigma_z  \left( \frac{|\ell|^2 w_0^2}{r^2} - 4 |\ell| + \frac{4 r^2}{w_0^2} - 2 \right) \right] \; , 
\end{equation}
where
\begin{equation}
    \psi_\ell = \frac{1}{\sqrt{|\ell|!}} \left( \frac{\sqrt{2} r}{w_0}\right)^{|\ell|} \left(\frac{x_R}{X} \right)^{|\ell| + 1} \exp \left( - \frac{k r^2}{2 X} \right) e^{i \ell \phi}
\end{equation}
is the azimuthal LG mode containing the Rayleigh range $x_R$ and the complex beam parameter $X = x_R + i x$, $r$ is the radial position and $w_0$ the beam waist. The parameter $\sigma_z$ determines the linear/circular polarization of the light. In the following, we will only consider linearly polarized pulses, i.e. $\sigma_z = 0$. While the peak intensities (and consequently, the strength of the generated field) for currently realizable Laguerre-Gaussian pulses is small compared to conventional Gaussian pulses, several methods for obtaining high-intensity LG pulses have been proposed \cite{Vieira2016, Zhang2025}.
Such pulses could be of interest for future application in ion acceleration, as their electromagnetic field structure has been shown to deliver well-collimated beams in several publications \cite{Hu2022, Dong2022, Wang2024}.

In this paper, we investigate the interaction of high-intensity LG laser pulses with targets of different hydrogen-boron ratio using 3D particle-in-cell (PIC) simulations. In particular, we focus on the generation of strong axial magnetic fields using the $\ell = 1, p = 0$ mode with near-critical density targets. Further, we study the interaction of these laser pulses with solid foils of different HB compositions.

\section*{Results}

\subsection*{Axial field generation in near-critical density HB compound targets}\label{sec:ncd}

We perform 3D particle-in-cell (PIC) simulations using the code \textsc{vlpl} \cite{Pukhov1999, Pukhov2016} with the rhombi-in-plane solver \cite{Pukhov2020}.
The simulation domain for near-critical density targets was chosen to be $100 \times 60 \times 60 \lambda_L^3$, with $\lambda_L = 800$ nm being the laser wavelength. The grid size is $h_x = 0.05 \lambda_L = c \Delta t$, $h_y = h_z = 0.1 \lambda_L$. From here on out, $x$ refers to the direction of laser propagation.
The incident laser pulse is a Laguerre-Gaussian pulse with $\ell = 1, p = 0$ and a normalized laser vector potential $a_0 = e E_L / m c \omega_L = 6.84$. Its focal spot size is $5 \lambda_L$ and its duration $10 \lambda_L/c$. At the beginning of the simulation ($t = 0$), the center of the laser pulse is located at $x = -20 \lambda_L$.

The composition of the near-critical density (NCD) targets is changed throughout the various simulation runs: the ratio of hydrogen to boron is varied in the range from 100:1 up to 1:100, however with a fixed total electron density of $0.02 n_\crit$.  
The initially unionized targets have a homogeneous density profile, and only have a small up-ramp of $5\lambda_L$ where the laser enters the target.

\begin{figure*}
    \centering
    \includegraphics[width=\textwidth]{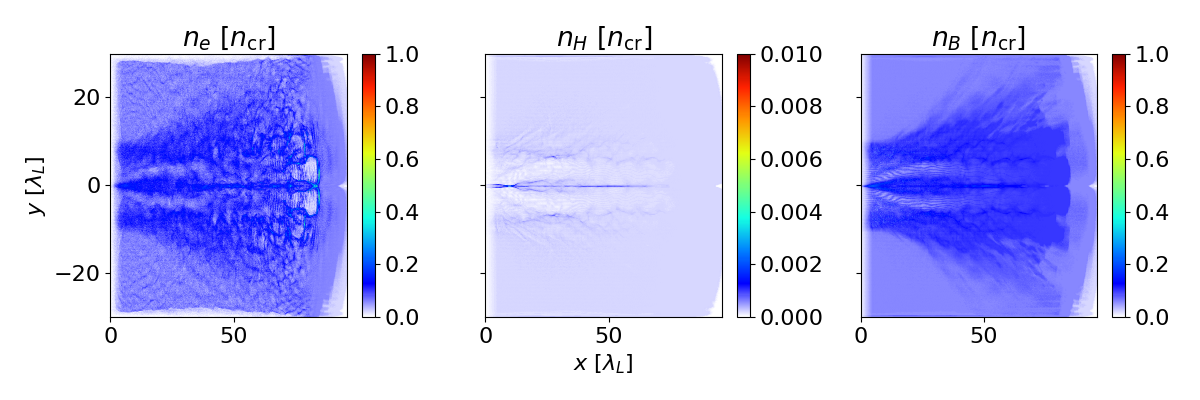}
    \caption{\label{fig:density}Densities for electrons (left), hydrogen ions (center) and boron ions (right) after $100 T_0$ for an HB ratio of 1:100. The laser pulse has ionized most of the hydrogen homogeneously and a well-defined channel with has formed. By contrast, only parts of the boron close to the laser position have been ionized to higher charge states, leading to a higher electron density around the laser region.}
\end{figure*}

The interaction of the LG pulse with a target of HB ratio 1:100 is exemplarily shown in figure \ref{fig:density}. When the laser pulse propagates through the target, its ponderomotive force pushes the electrons that have been field-ionized outwards from the high-intensity regions. 
Additionally, a dense ion filament (in the hydrogen component and -- to lesser extent -- in the boron component) is formed, which is similar to that in Magnetic Vortex Acceleration with two co-propagating pulses \cite{Reichwein2022}.
For an LG pulse, the electrons gain momentum in azimuthal direction, leading to the formation of a current $j_\phi$. According to Amper\`{e}'s law, this gives rise to an axial magnetic field $\partial_r (r B_x) / r = j_\phi$.

\begin{figure*}
    \centering
    \includegraphics[width=\textwidth]{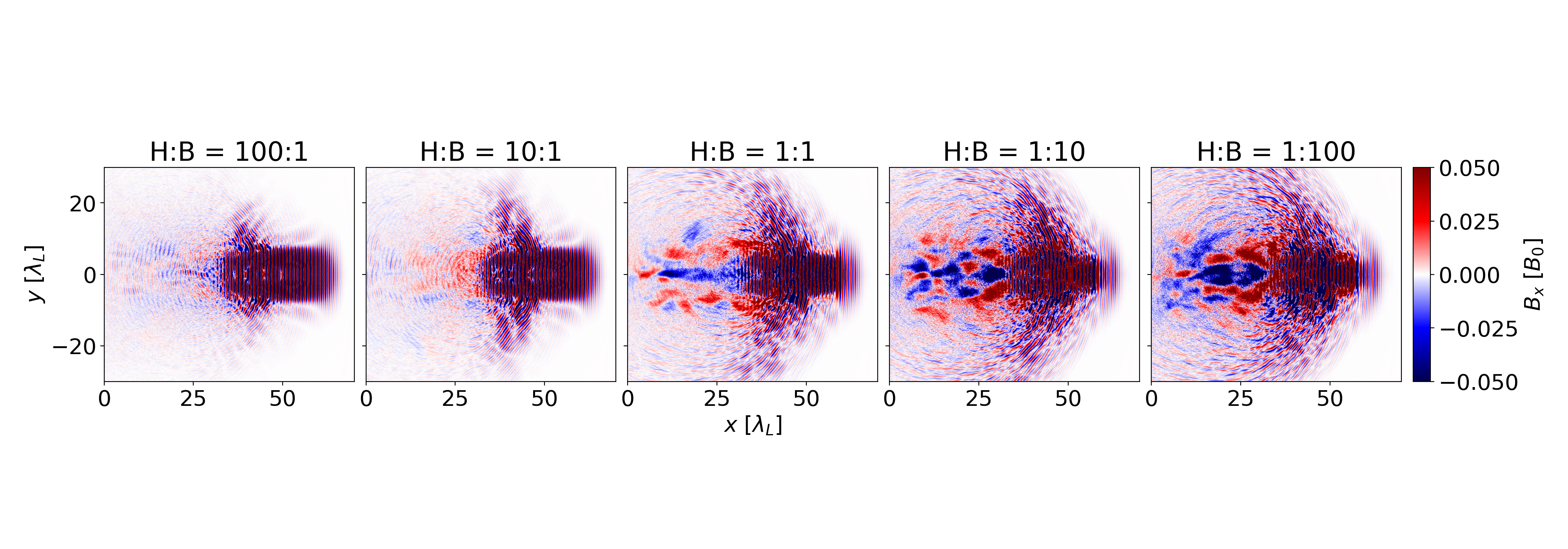}
    \caption{\label{fig:axial2d}Distribution of axial magnetic field $B_x$ after $70T_0$ for various HB ratios. Note that the colorbar is clipped at $\pm 0.05 B_0$ for better visibility.}
\end{figure*}

\begin{figure}
    \centering
    \includegraphics[width=0.5\textwidth]{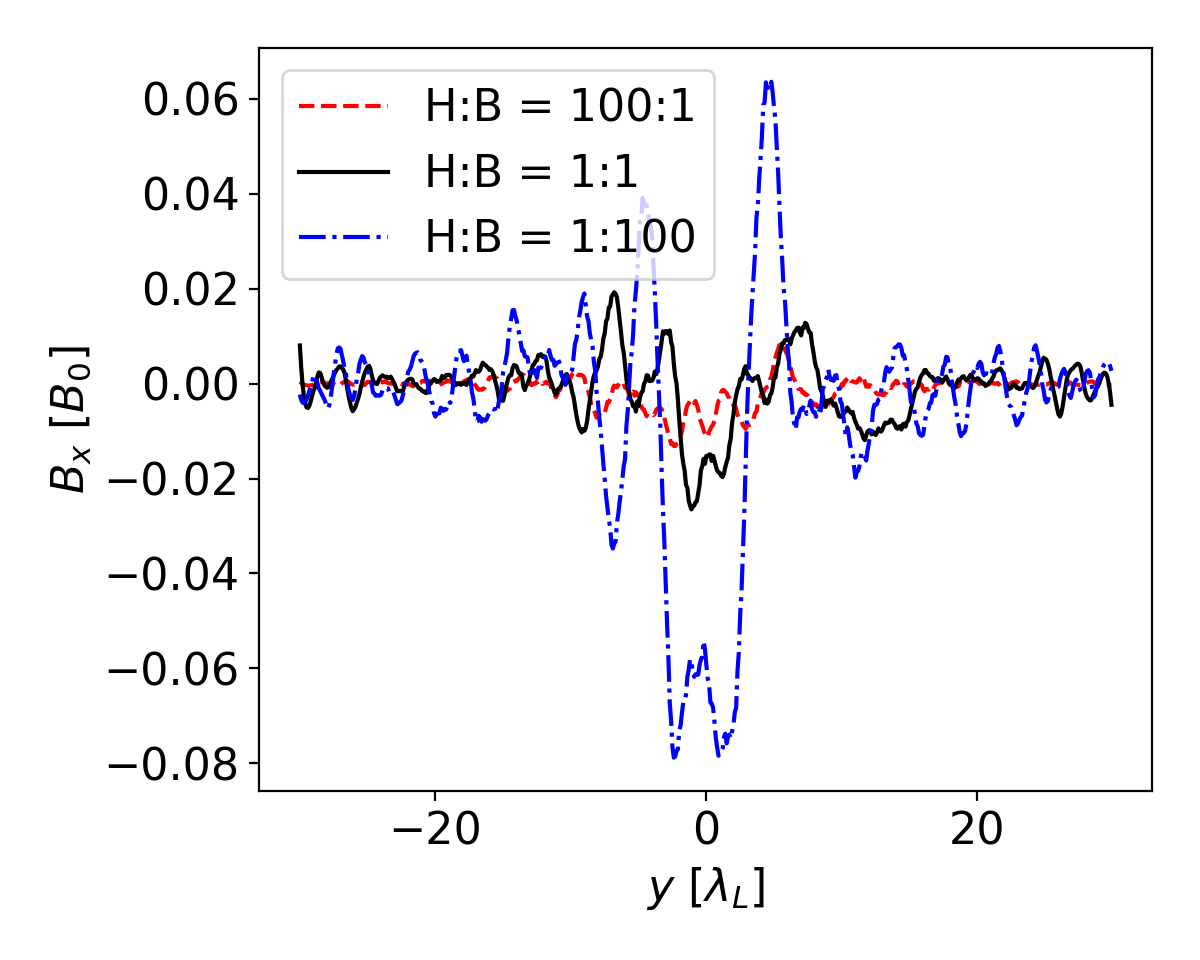}
    \caption{\label{fig:lineout}Transverse profile of the axial magnetic field structure for different HB ratios at $x = 15 \lambda$ after $70T_0$. A Savitsky-Golay filter has been utilized to denoise the data from the PIC simulations.}
\end{figure}

As shown in figure \ref{fig:axial2d}, the structure of the axial magnetic field is changing strongly depending on the HB ratio. In the case of a 1:1 ratio, a strong axial magnetic around $y = 0 \lambda$ is observed.
A smoothed transverse lineout of the axial fields is shown in figure \ref{fig:lineout}. We see that the axial fields are strongest for a higher boron percentage, while a higher hydrogen content leads to a reduction of the fields even below the case of a 1:1 composition.

One crucial difference here to prior studies like the ones by Longman and Fedosejevs \cite{Longman2021} on field generation by LG pulses is the fact, that we consider a completely unionized target. This means, that the amount of electrons available for angular momentum transfer is limited compared to a fully ionized target. While the difference in the first ionization level for the both components is not too large at $I_\mathrm{H} \approx 13.6$ eV and $I_\mathrm{B} \approx 8.3$ eV \cite{NIST}, it can make an important difference for ionization that occurs off-axis. Moreover, the HB ratio therefore determines the presence of an ionization front that is formed when the front of the laser pulse first irradiates the target, and that becomes important for subsequent momentum transfer once the highest-intensity part of the pulse reaches that region.

For higher borane content, the first ionization level is easily reached, but higher charge states have energy levels significantly beyond hydrogen (e.g. around 340 eV for B$^{5+}$). Thus, if not all boron atoms are fully ionized, boron-rich targets will have fewer electrons available for field generation. It has to be noted, that if the laser front already ionizes part of the target, these free electrons are available later for the main parts of the laser, since the laser essentially sees a pre-ionized target as it would in the case of Longman\cite{Longman2021}. This difference in ionization levels also is visible in figure \ref{fig:density}: around the area, where the laser pulses creates the plasma channel, a higher electron density can be seen. The additional electron stem from higher ionization levels of boron (see the darker blue area in the boron density), whereas the hydrogen component has already been fully ionized (visible as a homogeneous light blue in the density plot).

The increase in axial field strength in dependence of electron density is in agreement with expectations from the analytical theory of Longman\cite{Longman2021} which states that $B_x \propto \alpha_\mathrm{abs}/n_e$:
let us assume a target solely comprised of hydrogen that is irradiated by a laser pulse with just enough intensity to fully ionize it, yielding an effective electron density of $n_e$. By comparison, a target based on only boron would only be ionized to the state of $B^{1+}$, meaning that only $n_e/5$ would be available for field generation. In turn, the generated axial field should increase by a factor of five.
An increase in field strength for boron-rich targets is in fact observed, but the formation of this field is more complex due to the aforementioned reasons.
In physical reality, this simplified picture is more complicated due to different absorption coefficients for LG laser light which are generally not known (and not incorporated into PIC codes).

\begin{figure*}
    \centering
    \includegraphics[width=\textwidth]{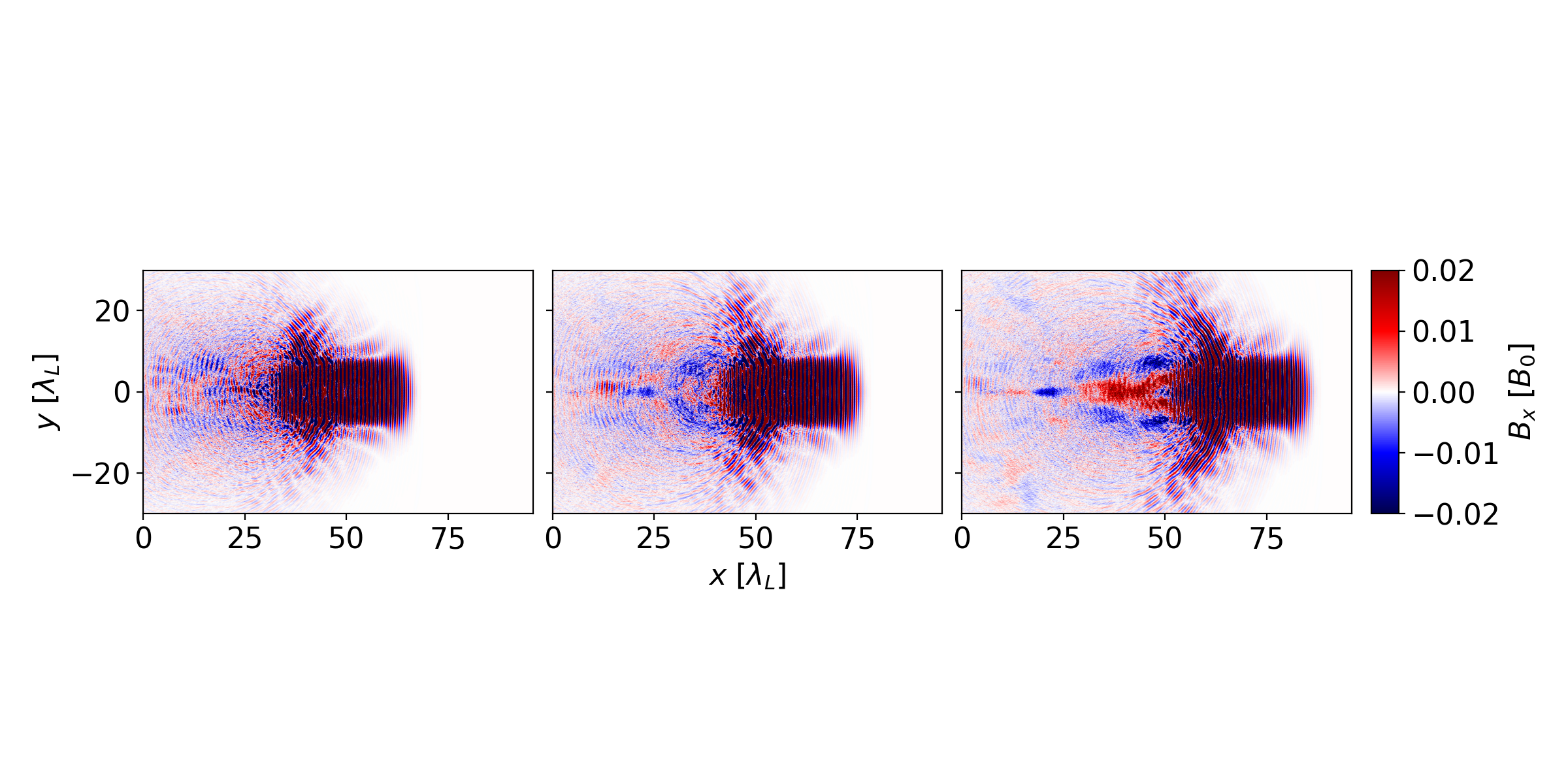}
    \caption{\label{fig:timesteps} Axial field generation for an HB ratio of 100:1 at different times $70 T_0$, $80 T_0$ and $90 T_0$ (left to right). The colorbar is clipped at $\pm0.02 B_0$ for better visibility. Initially, only a turbulent axial field structure is generated. Later on, a distinct forward-pointing axial field becomes visible.}
\end{figure*}

The complexity of axial field generation becomes even more apparent when taking into account several time steps. In figure \ref{fig:timesteps} we show the simulation with an HB ratio of 100:1 at different times. We observe that initially, only weaker and more turbulent axial fields are produced, while subsequently a dominant on-axis field of $+ 0.01 B_0$ is generated. This field is surrounded by a region of opposing polarity but similar field strength. By comparison (cf. figure \ref{fig:axial2d}), the simulations with a surplus of boron exhibit stronger fields already at initial stages ($70 T_0$), even if with different polarity.

The different ionization levels of the two target components also influence the plasma dynamics on longer timescales: the charge-to-mass ratio for hydrogen, $(q/m)_\mathrm{H} = 1$, is significantly larger than for boron, which ranges from $(q/m)_{\mathrm{B}^+} \approx 1/11$ to $(q/m)_{\mathrm{B}^{5+}} \approx 5/11$. Therefore, the expansion of the plasma will be significantly deviating for the boron-rich case when compared to hydrogen-rich targets.

\subsection*{Interaction of LG pulses with short foils}\label{sec:foils}
Beyond the investigation of underdense targets, we also study the interaction of Laguerre-Gaussian pulses with short, overdense foils. 
As before, the HB ratio of the targets is varied between different simulation runs, but the total electron density is fixed (here at $30n_\crit$). The foils are initially unionized and are modeled as 1 {\textmu}m thick slabs without any contaminant layers.

Since the geometry of the simulations has changed, we now employ a simulation domain of $50 \times 50 \times 50 \lambda_L^3$ and use the Yee-Maxwell solver with grid size $h_x = h_y = h_z = 0.05 \lambda_L$. The foil is located around the center of the box.
The laser pulse again has the mode $\ell = 1, p = 0$, now with a smaller focal spot of $3\lambda_L$ but same pulse duration of $10\lambda_L/c$. The intensity is increased to $a_0 = 30$.

\begin{figure*}
    \centering
    \includegraphics[width=\textwidth]{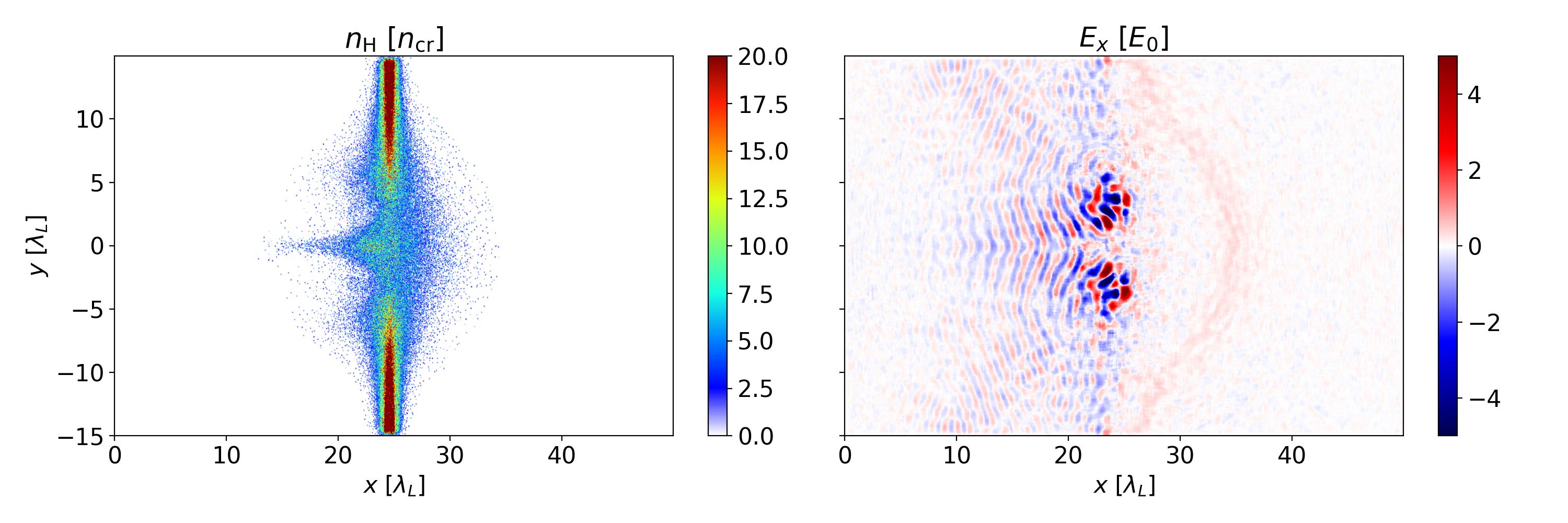}
    \caption{\label{fig:foil2d} Hydrogen density of a foil target with HB ratio of 100:1 after $100 T_0$ (left). The structure corresponds to the regions of peak intensity of the LG pulse. Right: Longitudinal electric field $E_x$. The color bars are clipped for better visibility.}
\end{figure*}

In the simulations, we observe that the laser pulse is able to extract most of the electrons of the thin foil in the interaction point. Due to the specific laser and target parameters, ions can are accelerated in parts due to Coulomb explosion of the target.
For an HB ratio of 100:1, the LG mode is imprinted in the Hydrogen density profile (cf. figure \ref{fig:foil2d}) as well as the electron density. In particular, a higher-density region is formed in the center of the LG pulse, where the intensity is zero. This feature is notably less pronounced/absent for  targets with higher boron content due to the change in $q/m$.
The laser intensity hot spots responsible for this structure are also visible in figure \ref{fig:foil2d}. Here, also the formation of a weak sheath field $E_x$ can be observed. This sheath field forms more slowly for boron-rich targets.

\begin{figure}
    \centering
    \includegraphics[width=0.5\textwidth]{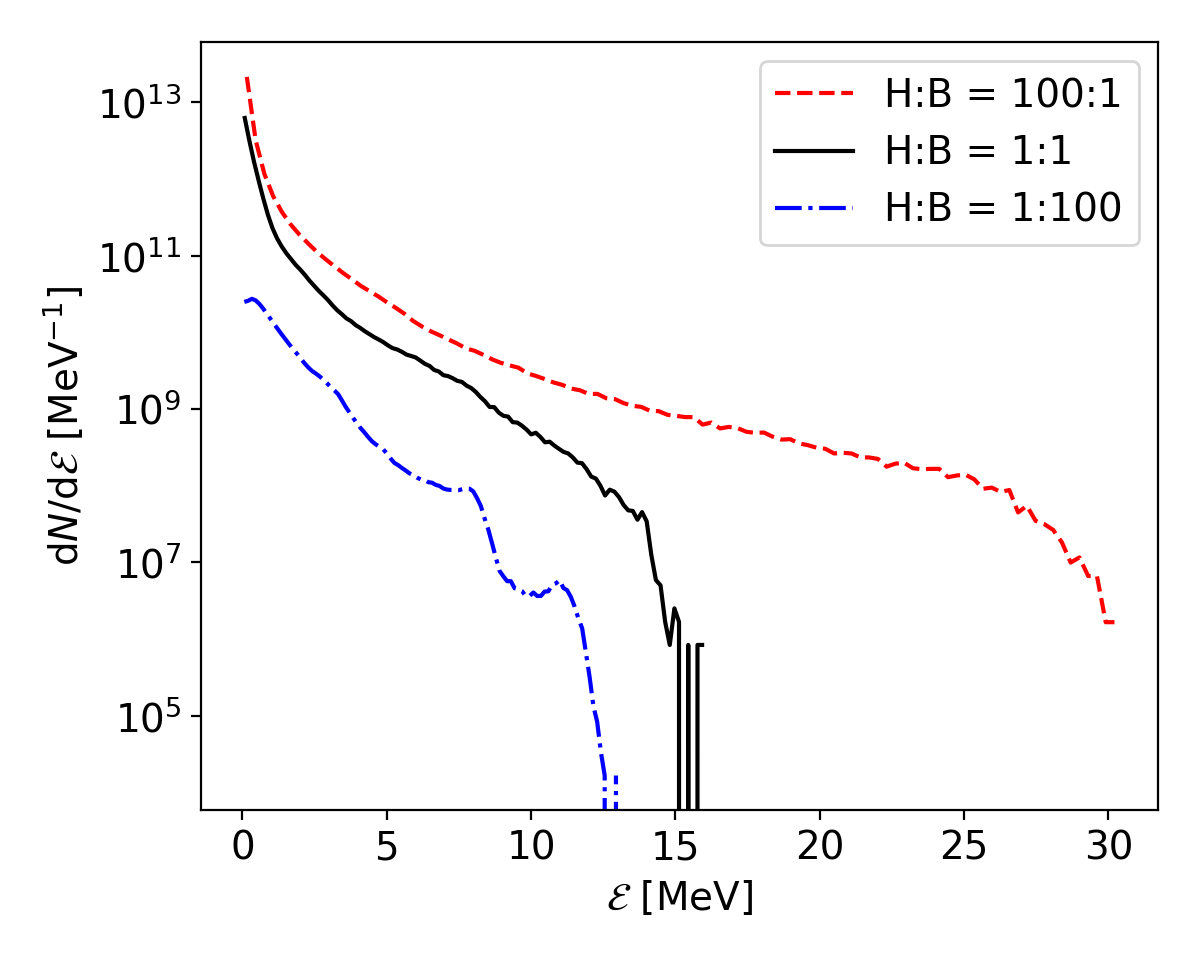}
    \caption{\label{fig:foil_spectrum} Proton energy spectra for different foil compositions after $160 T_0$. Hydrogen-rich compositions yield the highest energy and largest particle number.}
\end{figure}

Accordingly, hydrogen-rich targets deliver the highest proton energies at approximately 30 MeV for H:B = 100:1 (cf. figure \ref{fig:foil_spectrum}). Higher boron content leads to lower proton energies (approx. 12 MeV for H:B = 1:100) as well as an obvious decrease in the number of accelerated protons.
The energy spectra are thermal and no distinct peaks are observed in any of the simulations. The angular spread for simulations with higher boron percentage is slightly larger which can be attributed due to differences in electron heating by the laser pulse.

 \section*{Conclusion}
 In this paper, we have investigated the formation of axial magnetic fields using Laguerre-Gaussian laser pulses in underdense hydrogen-boron targets of different compound ratios. The axial fields are generated via the Inverse Faraday effect, as the laser pulse can transfer orbital angular momentum. We have found that a higher boron-content leads to the generation of stronger fields.
 This is due to the different charge states of hydrogen and boron, which yield effectively different electron densities for field generation.
 For solid foils, we observe that hydrogen-rich targets deliver the highest proton energies which can be of relevance for pitcher-catcher type setups.
 More generally, these results can be of particular interest for proton-boron fusion setups, where boron hydrides have recently been realized as possible fusion fuels \cite{Krus2024}.
 These compounds allow the specific ratio of hydrogen and boron to be tuned, changing the effective heating process and subsequent fusion reactions. 
 
 In a next step, the feasibility of different boranes for fusion should be verified experimentally, as first experiments show promising alpha-particle yields \cite{Krus2024}. 
 Accordingly, a next step from the theoretical research could be to investigate the interaction of longer laser pulses exceeding several 100 fs with realistic solid borane targets with respect to alpha-particle generation, which requires MHD simulations.
 Moreover, our current PIC-based study leaves out specific absorption coefficients. The exact absorption coefficients for laser light carrying OAM would be of interest for future studies: these effects have not been modeled in our current PIC simulations, but some research seems to indicate that OAM enhances absorption \cite{Peskosky2024}.

\section*{Data Availability}
The datasets generated during and/or analyzed during the current study are available from the corresponding
author on reasonable request.

%\bibliography{pB_ratio_bib.bib}

\section*{Acknowledgements}

This project has received funding from the European Union’s Horizon Europe research and innovation programme under grant agreement no 101096317.  UK participants in the Horizon Europe Project V4F are supported by UKRI grant number 10062154 (MODUS). 
The authors gratefully acknowledge the Gauss Centre for Supercomputing e.V. \cite{gcs} for funding this project (spaf) by providing computing time through the John von Neumann Institute for Computing (NIC) on the GCS Supercomputer JUWELS at Jülich Supercomputing Centre (JSC). The work of M.B. has been carried out in the framework of the JuSPARC (J\"{u}lich Short-Pulse Particle and Radiation Center \cite{jusparc}) and has been supported by the ATHENA (Accelerator Technology Helmholtz Infrastructure) consortium. 

\section*{Author contributions statement}

L.R. conducted the simulations and wrote the manuscript, with contributions from A.P. and M.B. All authors reviewed the manuscript. 

\section*{Additional information}

\textbf{Competing interests:} The authors declare no competing interest. 

\end{document}